\documentclass[twocolumn,showpacs,preprintnumbers,amsmath,amssymb,prl]{revtex4-1}%
\usepackage{graphicx}
\usepackage{dcolumn}
\usepackage{bm}
\usepackage{tabularx}
\usepackage{amsmath}
\usepackage{textcomp}
\usepackage{upgreek}

\begin{document}
\title{Doping evolution of the superconducting gap structure in heavily hole-doped Ba$_{1-x}$K$_x$Fe$_2$As$_2$: heat transport study}
\author{X. C. Hong,$^1$ A. F. Wang,$^2$ Z. Zhang,$^1$ J. Pan,$^1$ L. P. He,$^1$ X. G. Luo,$^2$ X. H. Chen,$^2$ and S. Y. Li$^{1,}$}
\email{shiyan$_$li@fudan.edu.cn}
\affiliation{$^1$State Key Laboratory of Surface Physics, Department of Physics, and Laboratory of Advanced Materials, Fudan University, Shanghai 200433, P. R. China\\ $^2$Hefei National Laboratory for Physical Science at Microscale and Department of Physics, University of Science and Technology of China, Hefei, Anhui 230026, P. R. China}

\date{\today}

\begin{abstract}
We performed systematic thermal conductivity measurements on heavily hole-doped Ba$_{1-x}$K$_x$Fe$_2$As$_2$ single crystals with 0.747 $\leq x \leq$ 0.974. At $x$ = 0.747, the $\kappa_0/T$ is negligible, indicating nodeless superconducting gap. A small residual linear term $\kappa_0/T$ ($\approx$ 0.035 mW/K$^2$ cm) appears at $x$ = 0.826, and it increases slowly up to $x$ = 0.974, followed by a drastic increase of more than 20 times to the pure KFe$_2$As$_2$ ($x$ = 1.0). This doping dependence of $\kappa_0/T$ clearly shows that the nodal gap appears near $x = 0.8$, likely associated with the change of Fermi surface topology. The small values of $\kappa_0/T$ from $x$ = 0.826 to 0.974 are consistent with the ``$\curlyvee$''-shaped nodal $s$-wave gap recently revealed by angle-resolved photoemission spectroscopy experiments at $x$ = 0.9. Furthermore, the drastic increase of $\kappa_0/T$ from $x$ = 0.974 to 1.0 is inconsistent with a symmetry-imposed $d$-wave gap in KFe$_2$As$_2$, and the possible nodal gap structure in KFe$_2$As$_2$ is discussed.

\end{abstract}

\pacs{74.70.Xa, 74.25.fc, 74.62.Dh}

\maketitle
To understand the pairing mechanism of electrons in superconductors, it is instructive to study the symmetry and structure of their superconducting gap. Due to the multiple Fermi surfaces (FSs), the situation in  iron-based high-temperature superconductors is very complicated \cite{Hirschfeld}.
While many of them have nodeless superconducting gaps, clear evidences for nodal gap were found in some iron-based superconductors, such as  KFe$_2$As$_2$ \cite{DongJK,Hashimoto-KFeAs,Reid,Okazaki}, LaFePO \cite{Fletcher,Hicks}, LiFeP \cite{Hashimoto-LiFeP}, BaFe$_2$(As$_{1-x}$P$_x$)$_2$ \cite{Nakai,Hashimoto-BaFeAsP,ZhangY}, and Ba(Fe$_{1-x}$Ru$_x$)$_2$As$_2$ \cite{QiuX}.
Clarifying whether the nodes are symmetry-imposed (like $d$-wave in cuprate superconductors) or just accidental is crucial for fully understanding the pairing mechanism in iron-based superconductors.

Among these nodal superconductors, one particular interesting case is KFe$_2$As$_2$ \cite{DongJK,Hashimoto-KFeAs,Reid,Okazaki,Tafti,Thomale}. Unlike the optimally doped Ba$_{0.6}$K$_{0.4}$Fe$_2$As$_2$ which has both electron and hole FSs \cite{DingH}, only hole FSs ($\Gamma$-centered pockets and off-$M$-centered lobes) were found in the extremely hole-doped KFe$_2$As$_2$ \cite{Sato}. Earlier thermal conductivity and penetration depth measurements unambiguously showed that KFe$_2$As$_2$ has nodal superconducting gap \cite{DongJK,Hashimoto-KFeAs}. Later, more detailed thermal conductivity study provided evidences for a $d$-wave gap \cite{Reid}. The recent observation of sudden reversal in the pressure dependence of $T_c$ was also interpreted as evidence for $d$-wave superconducting state in KFe$_2$As$_2$ at ambient pressure \cite{Tafti}. However, the laser angle-resolved photoemission spectroscopy (ARPES) experiments conducted at 2 K directly revealed the gap structure on the $\Gamma$-centered hole FSs: a nodeless gap on the inner FS, an unconventional gap with ``octet-line nodes'' on the middle FS, and an almost-zero gap on the outer FS, which showed that KFe$_2$As$_2$ is a nodal \emph{s}-wave superconductor \cite{Okazaki}.

To investigate the doping evolution of gap structure in heavily hole-doped Ba$_{1-x}$K$_x$Fe$_2$As$_2$ may shed light on this enigma.
Recently, laser ARPES measurements on Ba$_{1-x}$K$_x$Fe$_2$As$_2$ ($x$ = 0.93, 0.88, and 0.76) revealed systematic change of gap structure on the $\Gamma$-centered hole FSs upon Ba doping into KFe$_2$As$_2$ \cite{Ota}. While the inner FS remains nodeless, the nodes gradually disappear on the middle FS upon Ba doping, and eight nodes clearly appear on the outer FS for all three dopings \cite{Ota}. However, another ARPES group measured Ba$_{1-x}$K$_x$Fe$_2$As$_2$ with $x$ = 0.7 and 0.9 \cite{Nakayama,XuN}, and they did not find any nodes in all three $\Gamma$-centered hole FSs. While the gap in the $M$-centered electron pocket is nodeless for $x$ = 0.7, a ``$\curlyvee$''-shaped nodal gap was detected at the tip of the four off-$M$-centered small hole lobes for $x = 0.9$ \cite{Nakayama,XuN}. Therefore, there are two important issues of the superconducting gap in heavily hole-doped Ba$_{1-x}$K$_x$Fe$_2$As$_2$. One is at which doping level the superconducting gap changes from nodeless to nodal, and the other is what is the exact nodal gap structure in the heavily hole-doped regime. More experiments are highly desired to clarify these two related issues, which will also help to resolve the KFe$_2$As$_2$ enigma.

In this Letter, we present a systematic heat transport study of heavily hole-doped Ba$_{1-x}$K$_x$Fe$_2$As$_2$ single crystals with 0.747 $\leq x \leq$ 0.974.
According to the doping dependence of the residual linear term $\kappa_0/T$, the nodeless to nodal gap structure change happens near $x = 0.8$, likely coinciding with the Lifshitz transition (change of the FS topology). The small values of $\kappa_0/T$ from $x$ = 0.826 to 0.974 are consistent with the peculiar ``$\curlyvee$''-shaped nodal $s$-wave gap structure observed in Ref. \cite{XuN}. A more than 20 times increase of $\kappa_0/T$ from $x$ = 0.974 to $x$ = 1 is inconsistent with a symmetry-imposed $d$-wave gap in KFe$_2$As$_2$. The large value of $\kappa_0/T$ in KFe$_2$As$_2$ may be mainly caused by the drastic decrease of the slope of the gap at the node, i.e., changing from ``$\curlyvee$''-shaped nodal gap to some kind of ``$\vee$''-shaped.

\begin{figure}
\centering
\centering
\includegraphics[clip,width=0.40\textwidth]{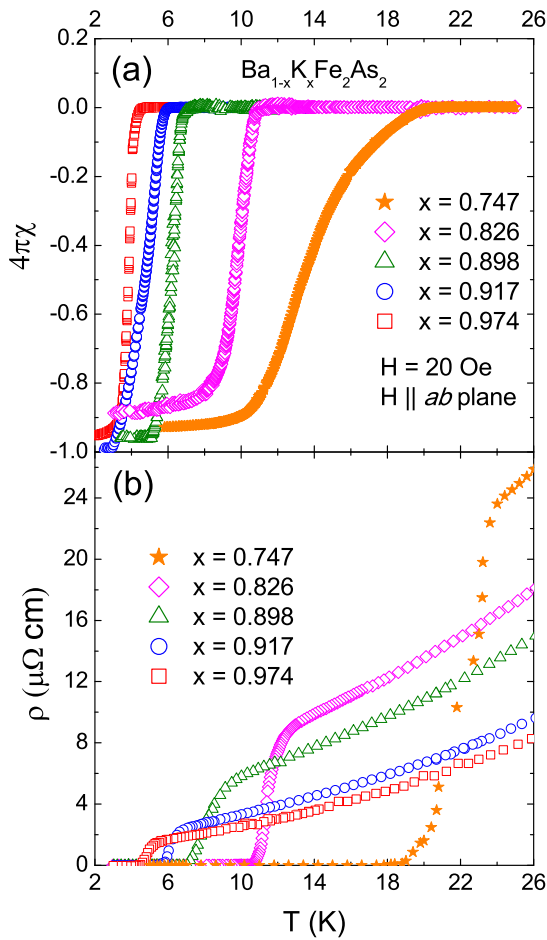}
\caption{(Color online) (a) The dc magnetic susceptibility of heavily hole-doped Ba$_{1-x}$K$_x$Fe$_2$As$_2$ single crystals with $x$ = 0.747, 0.826, 0.898, 0.917, and 0.974. (b) The low-temperature resistivity of these Ba$_{1-x}$K$_x$Fe$_2$As$_2$ single crystals. For each sample, the transition temperature $T_c$ defined by $\rho = 0$ agrees well with the onset of the diamagnetic transition in (a).}
\end{figure}

Single crystals of heavily hole-doped Ba$_{1-x}$K$_x$Fe$_2$As$_2$ were grown by the self-flux method \cite{WangAF}.
The compositions were determined by wavelength-dispersive spectroscopy (WDS), utilizing an electron probe micro-analyzer (Shimadzu EPMA-1720).
The dc magnetic susceptibility was measured at $H$ = 20 Oe, with zero-field cooling, using a SQUID (MPMS, Quantum Design).
The samples were cleaved and cut into rectangular shape with large \emph{ab} plane. Contacts were made directly on the sample surfaces with silver paint, which were used for both resistivity and thermal conductivity measurements. All samples were exposed in air less than 3 hours to avoid degradation.
The contacts are metallic with typical resistance 30 m$\Omega$ at 2 K. In-plane thermal conductivity was measured in a dilution refrigerator, using a standard four-wire steady-state method with \emph{in situ} calibrated RuO$_2$ chip thermometers.
Magnetic fields were applied along the \emph{c} axis and perpendicular to the heat current.
To ensure a homogeneous field distribution in the sample, all fields were applied at temperature above $T_c$.

\begin{table}
\centering \caption{The $T_c$, residual resistivity $\rho_0$, residual resistivity ratio RRR, residual linear term $\kappa_0/T$, the fitting parameter $\alpha$, and the normalized $\kappa_0/T$ of Ba$_{1-x}$K$_x$Fe$_2$As$_2$ single crystals with $x$ = 0.747, 0.826, 0.898, 0.917, and 0.974. The values of clean KFe$_2$As$_2$ single crystal are also listed \cite{Reid}.}\label{1}
\begin{tabularx}{0.48\textwidth}{p{0.9cm}p{0.9cm}p{1.2cm}p{0.8cm}p{2cm}p{0.8cm}p{1cm}p{0.1cm}}\hline\hline
$x$ & $T_c$ & $\rho_0$ & RRR & $\kappa_0/T$ & $\alpha$ & $\dfrac{\kappa_0/T}{\kappa_{N0}/T}$ &  \\
\\ & (K) & $(\mu\Omega$ cm) & & (mW/K$^2$ cm) & & & \\ \hline
0.747 & 18.7 &  17.9& 25     & 0     &2.05 &    0\% \\
0.826 & 10.8 &  6.15& 57     & 0.035 &3.34 & 0.88\% \\
0.898 & 7.16 &  2.83& 105    & 0.054 &3.09 & 0.62\% \\
0.917 & 5.66 &  1.38& 176    & 0.106 &3.15 & 0.60\% \\
0.974 & 4.55 &  0.76& 300    & 0.152 &3.76 & 0.48\% \\
1.0   & 3.80 &  0.21& 1108   & 3.60  &3    & 3.1\%  \\
\hline \hline
\end{tabularx}
\end{table}

Figure 1(a) shows the dc magnetic susceptibility of heavily hole-doped Ba$_{1-x}$K$_x$Fe$_2$As$_2$ single crystals with $x$ = 0.747, 0.826, 0.898, 0.917, and 0.974. The low-temperature resistivity of these Ba$_{1-x}$K$_x$Fe$_2$As$_2$ single crystals are plotted in Fig. 1(b). For each sample, the transition temperature $T_c$ defined by $\rho = 0$ agrees well with the onset of the diamagnetic transition in Fig. 1(a). Table I lists their $T_c$, residual resistivity $\rho_0$, and residual resistivity ratio RRR. The values of clean KFe$_2$As$_2$ single crystal are also listed \cite{Reid}.

Figure 2(a) presents the low-temperature thermal conductivity of the five heavily hole-doped Ba$_{1-x}$K$_x$Fe$_2$As$_2$ in zero field. To obtain the residual linear term $\kappa_0/T$, the curves are fitted to $\kappa/T$ = $\kappa_0/T + bT^{\alpha-1}$ below 0.25 K \cite{Sutherland,LiSY}. For $x$ = 0.747, the fitting gives $\kappa_0/T = -5 \pm 9$ $\mu$W K$^{-2}$ cm$^{-1}$. Comparing with our experimental error bar 5 $\mu$W K$^{-2}$ cm$^{-1}$, this value is negligible and we take it as zero. For 0.826 $\leq x \leq$ 0.974, the fittings gave finite but quite small $\kappa_0/T$, which are listed in Table I together with that of clean KFe$_2$As$_2$ sample \cite{Reid}. The second term $bT^{\alpha-1}$ is normally contributed by phonons, and the typical value of $\alpha$ lies between 2 and 3 \cite{Sutherland,LiSY}. However, as listed in Table I, the $\alpha$ values of 0.826 $\leq x \leq$ 0.974 samples are abnormally higher than 3, which will be discussed later.

In Fig. 2(b), we plot the doping dependence of $T_c$ and $\kappa_0/T$ for heavily hole-doped Ba$_{1-x}$K$_x$Fe$_2$As$_2$. While the $T_c$ shows a smooth doping dependence, the doping evolution of $\kappa_0/T$ is rather drastic \cite{Note}. The negligible $\kappa_0/T$ at $x$ = 0.747 suggests that it is still fully gapped, as the case of $x$ = 0.7 \cite{Nakayama}. A finite $\kappa_0/T$ appears at $x$ = 0.826, and increases slowly and monotonically up to $x$ = 0.974. For a superconductor, the finite $\kappa_0/T$ in zero field usually comes from the nodal quasiparticles \cite{Shakeripour}. Therefore, all these four samples have nodal superconducting gap. From $x$ = 0.974 to $x$ = 1, a more than 20 times increase of $\kappa_0/T$ is observed. Below we discuss this systematic doping evolution of $\kappa_0/T$ in Ba$_{1-x}$K$_x$Fe$_2$As$_2$.

\begin{figure}
\centering
\includegraphics[clip,width=0.41\textwidth]{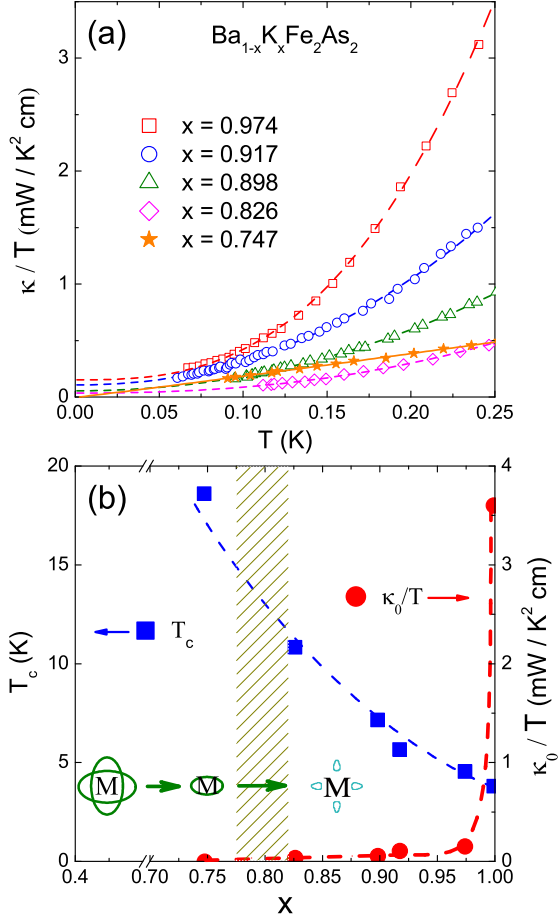}
\caption{(Color online) (a) Low-temperature in-plane thermal conductivity of Ba$_{1-x}$K$_x$Fe$_2$As$_2$ single crystals in zero field. The dashed lines are fitting curves according to $\kappa/T$ = $\kappa_0/T$ + $b$$T^{{\alpha}-1}$.
(b) Doping dependence of $T_c$ and $\kappa_0/T$.
The data of clean KFe$_2$As$_2$ single crystal are included \cite{Reid}.
The dashed lines are guide to the eye.
The Lifshitz transition is roughly located inside the shadow area \cite{XuN}.}
\end{figure}

\begin{figure}
\centering
\includegraphics[clip,width=0.495\textwidth]{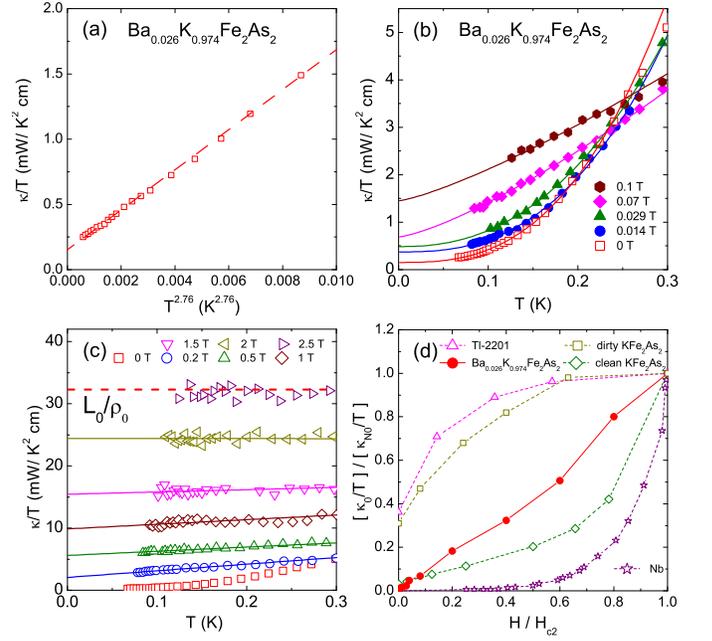}
\caption{(Color online) (a) Low-temperature in-plane thermal conductivity of Ba$_{0.026}$K$_{0.974}$Fe$_2$As$_2$ single crystal in zero field, plotted as $\kappa/T$ vs $T^{2.76}$. (b) and (c) $\kappa/T$ vs $T$ for Ba$_{0.026}$K$_{0.974}$Fe$_2$As$_2$ single crystal in magnetic fields applied along the $c$ axis. The solid lines are fits of the data to $\kappa/T$ = $a$ + $bT^{\alpha - 1}$. For $H$ $\geq$ 0.2 T, the curves are roughly linear, and $\alpha$ is fixed to 2. The dashed line is the normal-state Wiedemann-Franz law expectation $L_0/\rho_0$, with $L_0$ the Lorenz number 2.45 $\times$ 10$^{-8}$ W $\Omega$ K$^{-2}$ and $\rho_0$ = 0.76 $\mu\Omega$ cm. (d) Normalized residual linear term $\kappa_0/T$ of Ba$_{0.026}$K$_{0.974}$Fe$_2$As$_2$ as a function of $H/H_{c2}$. For comparison, similar data are shown for the clean $s$-wave superconductor Nb \cite{Lowell}, the archetypal $d$-wave cuprate superconductor Tl-2201 \cite{Proust}, the dirty and clean KFe$_2$As$_2$ \cite{Reid,DongJK}.}
\end{figure}

First, this doping evolution of $\kappa_0/T$ clarifies that the superconducting gap changes from nodeless to nodal between $x$ = 0.747 and 0.826. Previously, ARPES measurements show that there is a Lifshitz transition between $x$ = 0.7 and 0.9 \cite{Nakayama,XuN}. The $M$-centered small electron pockets at $x$ = 0.7 transforms into four off-$M$-centered hole lobes at $x$ = 0.9 \cite{Nakayama,XuN}. While the superconducing gaps in all FSs of $x$ = 0.7 sample are nodeless \cite{Nakayama}, there is a ``$\curlyvee$''-shaped vertical line node on each tip of the off-$M$-centered hole lobes for $x$ = 0.9 sample \cite{XuN}. Therefore, our thermal conductivity result is consistent with these ARPES experiments, and further narrows down the doping range to 0.747 $\leq$ $x$ $\leq$ 0.826, where the superconducting gap changes from nodeless to nodal. This change is very likely correlated to the Lifshitz transition \cite{Nakayama,XuN}, as illustrated in Fig. 2(b).

Second, we note that $\kappa_0/T$ is very small in the doping range of 0.826 $\leq$ $x$ $\leq$ 0.974 \cite{Note}. These values are only 0.48\% to 0.88\% of their normal-state Wiedemann-Franz law expectations $L_0/\rho_0$, with $L_0$ = 2.45 $\times$ 10$^{-8}$ W $\Omega$ K$^{-2}$. Theoretically, for a quasi-2D nodal superconductor, $\kappa_0/T$ can be estimated from \cite{Graf,MSuzuki}
\begin{eqnarray}
\frac{\kappa_0}{T} = \frac{\pi^2 k_B^2}{3}N_Fv_F^2\frac{a\hbar}{2\mu\Delta_0},
\end{eqnarray}
in which $N_F$ and $v_F$ are the density of states in the normal state and the Fermi velocity, respectively. The parameter $a$ depends on the gap symmetry, $\Delta_0$ represents the maximum amplitude of the gap, and $\mu$ is the slope of the gap at the node \cite{Graf,MSuzuki}. According to this formula, the small values of $\kappa_0/T$ for 0.826 $\leq$ $x$ $\leq$ 0.974 may be explained by the special nodal $s$-wave gap structure observed in Ba$_{0.1}$K$_{0.9}$Fe$_2$As$_2$ by ARPES \cite{XuN}. Since all of the $\Gamma$-centered hole FSs have nodeless gaps for $x$ = 0.9 \cite{XuN}, there is no contribution to $\kappa_0/T$ from these FSs. Nodes are only located at the tips of the off-$M$-centered $\varepsilon$ FS lobes \cite{XuN}. The angular distribution of the superconducting gap manifests a special ``$\curlyvee$'' shape near the node, as seen in Fig. 4(c) of Ref. \cite{XuN}. Comparing with the usual ``$\vee$'' shape near a $d$-wave gap node or the node observed in Ref. \cite{Ota}, the ``$\curlyvee$'' shape has a very large slop $\mu$ near the node, which may result in the very small $\kappa_0/T$ in the doping range of 0.826 $\leq$ $x$ $\leq$ 0.974.

Furthermore, the special ``$\curlyvee$''-shaped nodal $s$-wave gap structure may also explain the abnormally high fitting parameter $\alpha$ of the $bT^{\alpha}$ term in this doping range. Usually, the $bT^{\alpha}$ term is contributed by phonons, and $\alpha$ is between 2 and 3 \cite{Sutherland,LiSY}. However, the $\alpha$ values of 0.826 $\leq x \leq$ 0.974 samples are higher than 3. More interestingly, this $\alpha$ is very sensitive to magnetic field. As seen in Fig. 3(b), a very small field of $H$ = 0.1 T has suppressed $\alpha$ from 3.76 to 2.27 for $x$ = 0.974 sample. The high sensitivity of $\alpha$ to field suggests its electronic origin. In fact, for nodal superconductors, the $T$ dependence of thermodynamic quantities at low temperature depends on how the gap increases with the distance from the nodal point \cite{Matsuda}. For clean $d$-wave superconductors, an electronic $T^3$ term of thermal conductivity was theoretically predicted \cite{Graf}, and indeed observed in ultraclean YBa$_2$Cu$_3$O$_7$ single crystal \cite{Hill}. This electronic $T^3$ term was rapidly suppressed by magnetic field \cite{Hill}. In this context, the ``$\curlyvee$''-shaped nodal gap may result in an $\alpha$ value higher than 3, which needs to be theoretically verified. In Fig. 3(d), we plot the normalized $\kappa_0/T$ of Ba$_{0.026}$K$_{0.974}$Fe$_2$As$_2$ as a function of $H/H_{c2}$, obtained from Figs. 3(b) and 3(c). The curve lies between the dirty and clean KFe$_2$As$_2$, and is similar to those of RbFe$_2$As$_2$ and CsFe$_2$As$_2$ single crystals \cite{XCHong,ZhangZ}. For complex nodal $s$-wave gap structure of a multiband superconductor, with both nodal and nodeless gaps of different magnitudes, it is hard to get a theoretical curve of $\kappa_0(H)/T$.

Third, $\kappa_0/T$ increases more than 20 times from $x$ = 0.974 to $x$ = 1. This is hard to be simply explained by the decrease of impurity scattering, since $\rho_0$ decreases rather smoothly, only about 3 times form $x$ = 0.974 to $x$ = 1. Since the Fermi surface topology does not change from $x$ = 0.9 to $x$ = 1 \cite{Sato,XuN}, the superconducting pairing symmetry presumably remains unchanged at this doping regime. If this is the case, the drastic increase of $\kappa_0/T$ from $x$ = 0.974 to $x$ = 1 is against the $d$-wave pairing symmetry proposed in Ref. \cite{Reid}, since it is inconsistent with the universal heat conduction of a $d$-wave gap. In fact, according to Eq. (1), $\Delta_0$ is proportional to $T_c$, and $N_F$ and $v_F$ only change slightly in heavily hole-doped regime \cite{Malaeb,Hafiez}. The drastic increase of $\kappa_0/T$ must result from a drastic decrease of $\mu$. Therefore, while the decrease of impurity scattering may play some role, the drastic increase of $\kappa_0/T$ may be caused by the change of nodal gap from ``$\curlyvee$''-shaped to some kind of ``$\vee$''-shaped. In this scenario, the $\Gamma$-centered hole pockets remain nodeless in KFe$_2$As$_2$, which is in line with the gap structure of $x$ = 0.9 sample in Ref. \cite{XuN} and different from the ``octet-line nodes'' of $x$ = 1 sample in Ref. \cite{Okazaki}. To finally resolve this issue, more low-temperature ARPES measurements on high-quality KFe$_2$As$_2$ single crystals, other than Ref. \cite{Okazaki}, are needed.

In summary, we have measured the thermal conductivity of heavily hole-doped Ba$_{1-x}$K$_x$Fe$_2$As$_2$ single crystals with $x$ $\geq$ 0.747, and got a monotonic but drastic doping dependence of $\kappa_0/T$. The absence of $\kappa_0/T$ at $x$ = 0.747 and the appearance of a very small $\kappa_0/T$ ($\approx$ 0.035 mW K$^{-2}$ cm$^{-1}$) at $x = 0.826$ indicates that the crossover from nodeless to nodal superconducting state occurs at $x \sim 0.8$, correlated to the Lifshitz transition. The very small $\kappa_0/T$ and abnormally large $\alpha$ of the $x$ = 0.826, 0.898, 0.917 and 0.974 samples are consistent with the special ``$\curlyvee$''-shaped nodal $s$-wave gap structure observed in Ba$_{0.1}$K$_{0.9}$Fe$_2$As$_2$ \cite{XuN}. The more than 20 times jump of $\kappa_0/T$ from $x = 0.974$ to $x$ = 1.0 may be mainly caused by the change of nodal gap from ``$\curlyvee$''-shaped to some kind of ``$\vee$''-shaped on the off-$M$-centered hole lobes.

We thank T. X. Zhao for his help in WDS measurements and D. L. Feng for useful discussions. This work is supported by the Ministry of Science and Technology of China (National Basic Research Program No: 2012CB821402 and 2015CB921401), the Natural Science Foundation of China, Program for Professor of Special Appointment (Eastern Scholar) at Shanghai Institutions of Higher Learning, and STCSM of China (No. 15XD1500200).


\begin{thebibliography}{99}
\bibitem{Note} We note in a recent paper, the thermal conductivity of Ba$_{1-x}$K$_x$Fe$_2$As$_2$ single crystals with 0.76 $\leq x \leq$ 1 was measured down to $\sim$ 250 mK \cite{Watanabe}. Their results are very different from ours. They observed non-monotonic doping dependance of $\kappa_0/T$, and the $\kappa_0/T$ of their $x$ = 0.93 and 0.88 samples ($T_c$ = 7.3 and 11 K) are much larger than those of our samples with similar $T_c$.
\end{thebibliography}
\end{document}